\documentclass[prb,preprint]{revtex4-1} 

\usepackage{amsmath,amssymb,amsfonts}
\usepackage{cancel}
\usepackage{physics}
\usepackage{wasysym}
\usepackage{graphicx} 
\usepackage[tight,nice]{units}
\begin{document}


\title{Jacobi integral: A real-world application of the Lagrangian formulation}


\author{Jeremy A. Riousset}
\email{jeremy.riousset@erau.edu}
\affiliation{Department of Physical Sciences; Center for Space \& Atmospheric Research; Space and Atmospheric Instrumentation Laboratory, Embry-Riddle Aeronautical University, Daytona Beach, FL 32114}
\author{Manasvi Lingam}
\email{mlingam@fit.edu}
\affiliation{Department of Aerospace, Physics and Space Sciences, Florida Institute of Technology, Melbourne, FL 32901}
\affiliation{Department of Physics and Institute for Fusion Studies, The University of Texas at Austin, Austin, TX 78712}

\date{\today}

\begin{abstract}
 The applicability of advanced classical mechanics (viz., the Lagrangian and/or Hamiltonian approaches) to real-world problems may not always seem straightforward, despite the mathematical rigor and elegance of this field. Here, we present a proof of the Jacobi integral using the Lagrangian formulation as a viable alternative to the usual demonstration using Newton's second law. The result represents a useful example of how advanced classical mechanics can provide a significant advantage over standard methods (i.e., Newton's laws). We conclude with an illustration of the Jacobi integral in our Solar system: the Circular Restricted 3-Body Problem (CR3BP) around Pluto and Charon.
\end{abstract}

\maketitle 

\section{Introduction}\label{sec:Intro}

Courses in advanced classical mechanics, which are typically oriented toward covering Lagrangian and Hamiltonian formulations, are ubiquitous at the (under)graduate level for physics and astronomy majors, among others. In parallel, a vast number of textbooks have been devoted to these topics \citep{VIA89,DTG97,JS98,Goldstein:2002,WG03,KB04,PVP05,DM08,DS09,ED10,JA12,SW15,TB17}. Many, although by no means all, of these textbooks focus on developing the mathematical machinery underpinning Lagrangians and Hamiltonians, thus devoting comparatively less space to real-world applications of these mathematically elegant formulations; in actuality, the practical applications range from celestial mechanics (addressed below) to fluids and plasmas \citep{ZMR85,RS88,HMR98,PJM98,ML15,PJM17}. 

One of the most common real-world applications of such classical mechanics is in the realm of celestial mechanics, orbital mechanics, and astrodynamics \citep{BC61,FRM70,GND75,SMK95,SM98,MD99,RHB99,RF12,Vallado:2013,KFW15,GS16,Curtis:2020,BMW20,GRH22,ST23}. The majority of textbooks in these disciplines include an exposition of the Circular Restricted 3-Body Problem (CR3BP) and the Jacobi integral \citep{VS67,CM90,BG97,MV05}, owing to their widespread relevance and utility in astronomy, planetary science, and aerospace engineering; the 3-body problem has a long and distinguished history, as chronicled in Refs.~\onlinecite{BG97,MCG98,VAK16}.

Standard textbook treatments of the aforementioned subjects have used Newton's laws to derive the Jacobi integral (e.g., Refs.~\onlinecite{MD99,RHB99,Vallado:2013,DePater:2015,KFW15,GS16,Curtis:2020}). However, the derivation can become tedious, and the same results may be elegantly achieved via the Lagrangian formalism, thereby offering an ideal example of a real-world situation wherein advanced techniques of classical mechanics have an edge over elementary tools (Newton's laws) and simplify the derivations, while offering mathematical and physical insights into the process. Chapter 10 of Ref.~\onlinecite{VS67} presents an explicit Lagrangian perspective on the Jacobi integral, but this specialized monograph is not readily accessible to (under)graduate physics audiences, and the exposition is not compact and entirely self-contained. Furthermore, no specific real-world examples of this concept were worked out in Ref.~\onlinecite{VS67}.

In this paper, we present a concise and self-contained derivation of the Jacobi constant -- also known as the Jacobi integral -- from the standpoint of Lagrangian mechanics in Section \ref{sec:Methods}, which can be of pedagogical value in teaching courses in advanced classical mechanics or cognate fields. We follow this derivation up with an illustration of the CR3BP and Jacobi integral in Sections \ref{sec:Results} and \ref{sec:Discussion} by focusing on the Pluto--Charon system. Finally, we summarize our salient results in Section \ref{sec:Conclusions}. 

\section{Jacobi constant (integral) derivation}\label{sec:Methods}
Hereafter, we will work with specific energy (i.e., energy per unit mass) and therefore will omit the term ``specific'' for the sake of brevity. First, let us define a non-inertial reference frame associated with the masses $m_1$ and $m_2$. The $x$-axis is given by the direction from $m_1$ to $m_2$, and the $z$-axis is perpendicular to the orbital plane of $m_1$ and $m_2$. The $y$-direction results from the cross-product $\vu*e_y=\vu*e_z\cp\vu*e_x$. It is worth recalling that in the system of uniformly rotating coordinates $(Oxyz)$, the masses $m_1$ and $m_2$ are at rest.

\begin{figure}[!ht]
  \centering
  \includegraphics[width=4in]{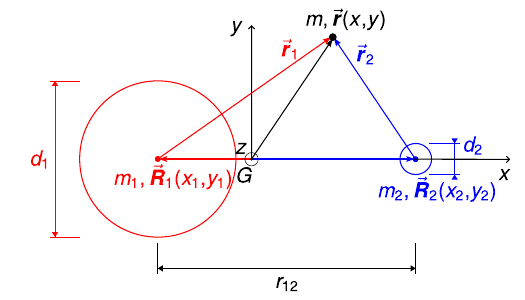}
  \caption{The Circular Restricted 3-Body Problem (CR3BP). Two point-masses $m_1$ and $m_2$ are located at the center of two spheres of diameters $d_1$ and $d_2$, placed at $\va*R_1$ and $\va*R_2$. They are at a distance of $r_{12}$ from each other. The third mass $m$ is located at $\va*r$ from the center of mass $G$, and at $\va*r_1$ and $\va*r_2$ from $m_1$ and $m_2$, respectively.}
  \label{fig:Setup}
\end{figure}

The third body of mass $m$, located at $\va*r$ has coordinates $(x,y)$ (Fig.~\ref{fig:Setup}). In a CR3BP, the center of mass $G$ is at rest ($\va*v_G=\va*0$) and the two-mass system comprising $m_1$ and $m_2$ is in uniform angular rotation, so that $\va*\Omega=\dfrac{2\pi}{P}\vu*e_z$, where $P$ denotes the orbital period that is given by Kepler's third law:
\begin{align}
  P & = \sqrt{\dfrac{4\pi^2}{\mu}r_{12}^3}\qq{where $r_{12}=\norm{\va*R_2-\va*R_1}$}.
\end{align}
Thus, the relative position, velocity, and acceleration can be expressed as follows:
\begin{align}
  \left\{
  \begin{array}{rcl}
    \va*{r}_{\rm rel} = \va*r & = & x\vu*e_x + y\vu*e_y               \\
    \va*{v}_{\rm rel} = \va*v & = & \dot{x}\vu*e_x + \dot{y}\vu*e_y   \\
    \va*{a}_{\rm rel} = \va*a & = & \ddot{x}\vu*e_x + \ddot{y}\vu*e_y \\
    \va*\Omega                & = & \Omega\vu*e_z
  \end{array}
  \right..\label{eq:SV}
\end{align}

Let us denote $\va*v_{\rm abs}$ the absolute velocity in the inertial reference frame. Under the aforementioned assumptions, the three-term velocity equation yields:
\begin{align}
  \va*{v}_{\rm abs} & = \cancelto{\va*0}{\va*{v}_G} + \va*{v}_{\rm rel} +  \va*{\Omega}\times \va*{r}_{\rm rel}\nonumber \\
                    & = (\dot{x}-\Omega y)\vu*e_x + (\dot{y}+\Omega x)\vu*e_y.
\end{align}
Consequently, the specific kinetic energy is:
\begin{align}
  T & =\dfrac{\va*v_{\rm abs}^2}{2}=\dfrac{\left(\dot{x}-\Omega y\right)^2+\left(\dot{y}+\Omega x\right)^2}{2}.\label{eq:T}
\end{align}


Next, let us denote by $\va*r_1$ and $\va*r_2$ the relative positions of $m$ with respect to $m_1$ and $m_2$ (Fig.~\ref{fig:Setup}). We can straightforwardly write:
\begin{align}
  \left\{
  \begin{array}{rcl}
    \va*{r}_1 & = & \va*r-\va*R_1 = (x-x_1)\vu*e_x +(y-y_1)\vu*e_y \\
    \va*{r}_2 & = & \va*r-\va*R_2 = (x-x_2)\vu*e_x +(y-y_2)\vu*e_y
  \end{array}\label{eq:r12}.
  \right.
\end{align}

The gravitational potential $V$ (or specific gravitational potential energy) is defined as:
\begin{align}
  V = \sum_{i\in\{1,2\}}\int_r^{\infty}\dfrac{\va*F_i}{m}\cdot\dd{\va*r}_i=-\dfrac{\mu_1}{r_1}-\dfrac{\mu_2}{r_2}.\label{eq:V}
\end{align}

The Lagrangian is conventionally defined as\cite{Goldstein:2002}:
\begin{align}
  L & =T-V.
  \label{eq:L}
\end{align}
Substituting \eqref{eq:T} and \eqref{eq:V} into \eqref{eq:L} lets us write:
\begin{align}
  L & =\dfrac{\left(\dot{x}-\Omega y\right)^2+\left(\dot{y}+\Omega x\right)^2}{2}+\dfrac{\mu_1}{r_1}+\dfrac{\mu_2}{r_2}.
  \label{eq:Lxy}
\end{align}
Eq. \eqref{eq:Lxy} lets us apply Lagrange's equations to obtain the equations of motion (Ref.~\onlinecite[p.~21]{Goldstein:2002}):
\begin{align}
  \dv{}{t}\pdv{L}{\dot{q}_i} & =\pdv{L}{q_i}\qq{where $\left\{\begin{array}{rclcrcl}q_1&=&x&;& \dot{q}_1&=&\dot{x}\\q_2&=&y&;&\dot{q}_2&=&\dot{y}\end{array}\right.$},\label{eq:DqL}
\end{align}
which can be understood as a mathematical formulation of the stationary-action principle or Hamilton's principle of least action \citep{VIA89,Goldstein:2002,WG03,DS09,ED10,TB17}. Starting with $i=1$, the left-hand side (LHS) of \eqref{eq:DqL} gives:
\begin{subequations}
  \begin{align}
    \dv{}{t}\pdv{L}{\dot{q}_1} & =  \dv{}{t}\pdv{}{\dot{x}}\left(\dfrac{\left(\dot{x}-\Omega y\right)^2+\left(\dot{y}+\Omega x\right)^2}{2}+\dfrac{\mu_1}{r_1}+\dfrac{\mu_2}{r_2}\right)=\dv{(\dot{x}-\Omega y)}{t}=\ddot{x}-\Omega\dot{y}\label{eq:LHS1}
  \end{align}
and similarly,
  \begin{align}
    \dv{}{t}\pdv{L}{\dot{q}_2} & =  \ddot{y}+\Omega\dot{x}\label{eq:LHS2}.
  \end{align}
\end{subequations}
On the other hand, the right-hand side (RHS) of \eqref{eq:DqL} gives:
\begin{align*}
  \pdv{L}{q_1} & =  \pdv{}{x}\left(\dfrac{\left(\dot{x}-\Omega y\right)^2+\left(\dot{y}+\Omega x\right)^2}{2}+\dfrac{\mu_1}{r_1}+\dfrac{\mu_2}{r_2}\right) =   \Omega(\dot{y}+\Omega x)     -\dfrac{\mu_1}{r_1^2}\pdv{r_1}{x}   -\dfrac{\mu_2}{r_2^2}\pdv{r_2}{x}.
\end{align*}
By employing the relations $\displaystyle\pdv{r_1}{x}=\dfrac{x-x_1}{r_1}$ and $\displaystyle\pdv{r_2}{x}=\dfrac{x-x_2}{r_2}$ from \eqref{eq:r12}, we have:
\begin{subequations}
  \begin{align}
    \pdv{L}{q_1} & = \Omega\dot{y}  +\Omega^2x -\dfrac{\mu_1(x-x_1)}{r_1^3} -\dfrac{\mu_2(x-x_2)}{r_2^3}.\label{eq:RHS1}
  \end{align}
Similarly, we can write $\displaystyle\pdv{r_1}{y}=\dfrac{y-y_1}{r_1}$ and $\displaystyle\pdv{r_2}{y}=\dfrac{y-y_2}{r_2}$ to obtain:
  \begin{align}
    \pdv{L}{q_2} & =  -\Omega\dot{x}    +\Omega^2y -\dfrac{\mu_1(y-y_1)}{r_1^3} -\dfrac{\mu_2(y-y_2)}{r_2^3}.\label{eq:RHS2}
  \end{align}
\end{subequations}
Setting \eqref{eq:LHS1} equal to \eqref{eq:RHS1} and \eqref{eq:LHS2}  equal to \eqref{eq:RHS2} leads to the result:
\begin{subequations}
  \begin{align}
    \ddot{x} & =  2n\dot{y} +\Omega^2x -\dfrac{\mu_1(x-x_1)}{r_1^3} -\dfrac{\mu_2(x-x_2)}{r_2^3} \label{eq:DDx}, \\
    \ddot{y} & = -2n\dot{x} +\Omega^2y -\dfrac{\mu_1(y-y_1)}{r_1^3} -\dfrac{\mu_2(y-y_2)}{r_2^3} \label{eq:DDy}.
  \end{align}
\end{subequations}
Finally, we calculate $\ddot{x}\dot{x}+\ddot{y}\dot{y}$ and simplify to get:
\begin{align}
  \ddot{x}\dot{x} + \ddot{y}\dot{y} & =\Omega^2(x\dot{x}+y\dot{y})  - \mu_1\dfrac{(x-x_1)\dot{x}+(y-y_1)\dot{y}}{r_1^3} - \mu_2\dfrac{(x-x_2)\dot{x}+(y-y_2)\dot{y}}{r_2^3}.\label{eq:Dv2}
\end{align}
The application of the chain rule to $\displaystyle\dv{}{t}\dfrac{1}{r_1}$ yields:
\begin{subequations}
  \begin{align}
    \dv{}{t}\dfrac{1}{r_1} & =\pdv{}{x}\dfrac{1}{r_1}\cdot\pdv{x}{t}+\pdv{}{y}\dfrac{1}{r_1}\cdot\pdv{y}{t} =-\dfrac{1}{r_1^2}\dfrac{x-x_1}{r_1}\dot{x}-\dfrac{1}{r_1^2}\dfrac{y-y_1}{r_1}\dot{y}\label{eq:inv1}.
  \end{align}
  The same approach for $\displaystyle\dv{}{t}\dfrac{1}{r_2}$ straightforwardly returns:
  \begin{align}
    \dv{}{t}\dfrac{1}{r_2} & =-\dfrac{(x-x_2)\dot{x}+(y-y_2)\dot{y}}{r_2^3} \label{eq:inv2}.
  \end{align}
\end{subequations}
Substituting \eqref{eq:inv1} and \eqref{eq:inv2} into \eqref{eq:Dv2} reduces the latter to:
\begin{align*}
  \ddot{x}\dot{x} + \ddot{y}\dot{y} & = \Omega^2(x\dot{x}+y\dot{y}) + \mu_1\dv{}{t}\dfrac{1}{r_1} +\mu_2\dv{}{t}\dfrac{1}{r_2}                                                  \\
  \Rightarrow 0                     & =  \dv{}{t}\left( \dfrac{\Omega^2\left(x^2+y^2\right)}{2} + \dfrac{\mu_1}{r_1} +\dfrac{\mu_2}{r_2} -\dfrac{\dot{x}^2+\dot{y}^2}{2}\right).
\end{align*}
This last equation demonstrates that the quantity inside the parenthesis is constant. Using \eqref{eq:SV}, we note that: $r^2=x^2+y^2$ and $v^2=\dot{x}^2+\dot{y}^2$ and thus we obtain:
\begin{align}
  \boxed{C_{\rm J}= \Omega^2r^2 + \dfrac{2\mu_1}{r_1} + \dfrac{2\mu_2}{r_2} - v^2},\label{eq:CJ}
\end{align}
where $C_{\rm J}$ is the Jacobi constant, also known as the Jacobi integral. Various versions of this equation can be found in the literature with alternative definitions of the constant (e.g., Ref.~\onlinecite[p.~127]{Curtis:2020}) or use of normalized units (e.g., Ref.~\onlinecite[p.~27]{DePater:2015} and Ref.~\onlinecite[p.~970]{Vallado:2013}). However, their proofs are usually lengthy and intricate, because they entail the application of Newton’s second law with appropriate geometric considerations. In contrast, the above derivation based on the Lagrangian formalism simplifies the proof, and constitutes an excellent example of the Lagrangian formulation evincing a substantial edge over Newtonian mechanics, for what turns out to be a classic textbook topic in orbital mechanics. 

Another pathway toward establishing the nature of the Jacobi integral, which we do not explicitly work out here, is to first construct the Hamiltonian $H$ for the CR3BP from the Lagrangian $L$, given by \eqref{eq:Lxy}, via the Legendre transform \citep{VIA89,Goldstein:2002,WG03,TB17}. Next, it can be demonstrated that $\dv*{C_{\rm J}}{t} \equiv \pb{C_{\rm J}}{H} = 0$ for this system, where $\pb{\cdot}{\cdot}$ is the canonical Poisson bracket, thereby confirming that the Jacobi integral is a constant of motion.

In the forthcoming section, we further develop a specific example of applying the Jacobi constant \eqref{eq:CJ} to an object in orbit around the Pluto (P)--Charon (C) system.

\section{Results: The Pluto--Charon example}\label{sec:Results}
\begin{table}[!b]
  \centering
  \begin{tabular}{lcccc}
    \hline\hline
    Object & Symbol  & Mass                & Radius      & Distance $Gm_i$ \\
           &    & \unit{(kg)}         & \unit{(km)} & \unit{(km)}     \\
    \hline
    Pluto  & P & $1.31\times10^{22}$ & 1188.3      & 2122.4          \\
    Charon & C & $1.59\times10^{21}$ & 606.0       & 17518.0         \\
    \hline\hline
  \end{tabular}
  \caption{Data sheet for the Pluto--Charon system.}
  \label{tab:PC_data}
\end{table}

The Pluto (P)--Charon (C) system provides a particularly telling example of the physical meaning of~\eqref{eq:CJ}, as demonstrated hereafter. This emphasis on a single system serves to focus attention on a concrete real-world scenario, and the advantages of the latter pedagogical strategy are well documented for students and educators alike \citep{JB93,MM93,CL96,LDMM,BAT03,DP05,RT07,JG08,RH08,King2012,RS17,BK22}. Owing to the large mass of Charon with respect to Pluto ($m_{\rm C}\approx0.12m_{\rm P}$) and relative proximity ($d_{\rm P-C}\approx16.5R_{\rm P}$), a third body in orbit around this system will experience the gravitational fields of both objects providing an ideal case study of CR3BP in our Solar system.

\begin{figure}[!ht]
  \centering
  \includegraphics[width=3in]{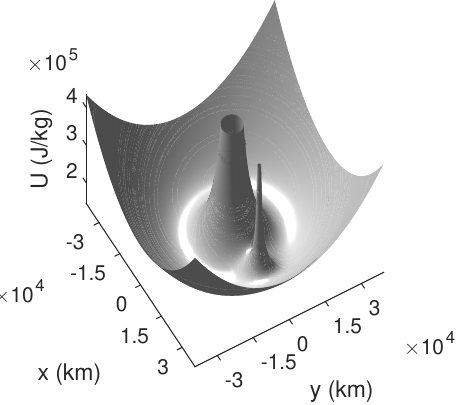}
  \caption{Pseudo potential: $U=C_{\rm J}+v^2=\Omega^2r^2 + \dfrac{2\mu_1}{r_1} + \dfrac{2\mu_2}{r_2}$ vs. $x-y$ (see Fig.~\ref{fig:Setup} and Ref.~\onlinecite{MatlabMonkey:2023}).}
  \label{fig:PP}
\end{figure}

Using the parameters summarized in Tab.~\ref{tab:PC_data}, Fig.~\ref{fig:PP} displays the pseudo-potential $U$ in the system of coordinates shown in Fig.~\ref{fig:Setup} with $U$ defined as follows:
\begin{align}
     U=C_{\rm J}+v^2=\Omega^2r^2 + \dfrac{2\mu_1}{r_1} + \dfrac{2\mu_2}{r_2}.\label{eq:PP}
\end{align}
This quantity resembles a pseudo-potential in the sense of combining rotational kinetic energy (or potential of the centrifugal pseudo-force) and gravitational potential. At the center of the largest peak and closest to the center of mass $G(0,0)$ lies Pluto, whereas Charon is responsible for the second peak in the vicinity of $(0,\unit[17500]{km})$.

\begin{figure}[!ht]
  \centering
  \includegraphics[width=\textwidth]{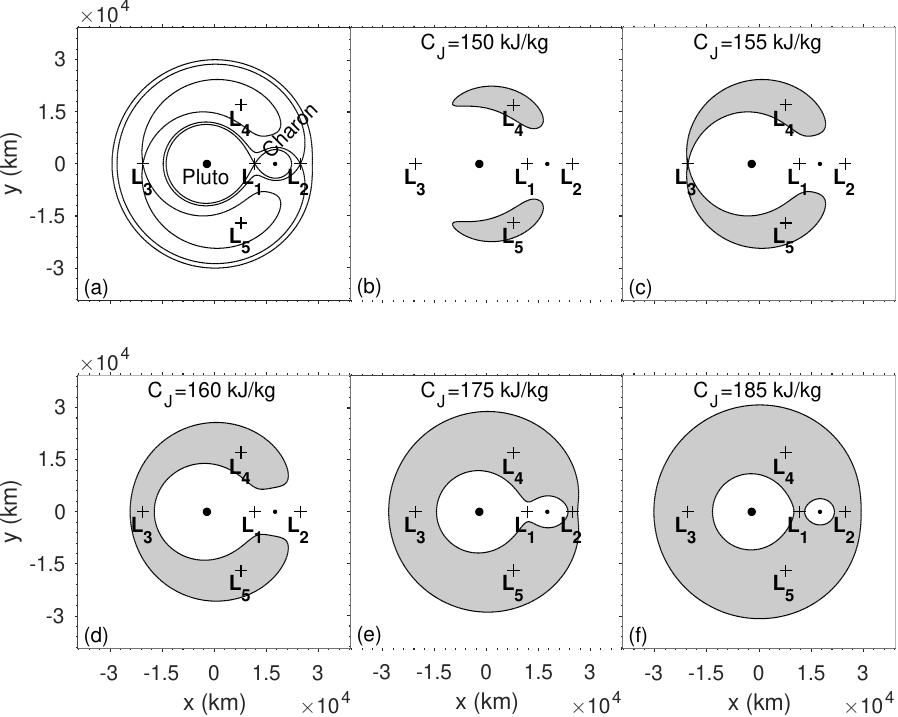}
  \caption{Zero-velocity surfaces for the  Pluto--Charon (P--C) system for multiple $C_{\rm J}$ constants. Adapted from Ref.~\onlinecite{MatlabMonkey:2023}.}
  \label{fig:JI}
\end{figure}

On the other hand, Fig.~\ref{fig:JI} displays the zero-velocity curves (a) and zero-velocity surfaces (b-f) for the following value of the Jacobi constant (or Jacobi integral): $C_{\rm J}\in\unit[\{150, 155, 160, 175, 185\}]{kJ/kg}$.

\section{Discussion}\label{sec:Discussion}
The rotating, non-inertial nature of the CR3BP reference frame does not enforce the conservation of energy and angular momentum. However, for the sake of clarity, we can rewrite \eqref{eq:PP} in the following fashion:

\begin{align}
    \underbrace{\left(\dfrac{1}{2}v^2-\dfrac{1}{2}\Omega^2r^2\right)}_{\mbox{(I)}}+\underbrace{\left(-\dfrac{\mu_1}{r_1}-\dfrac{\mu_2}{r_2}\right)}_{\mbox{(II)}} &=  -\dfrac{1}{2}C_{\rm J} = \dfrac{\mathcal{E}}{m}, \label{eq:Energy}
\end{align}

which represents the preferred form in Ref.~\onlinecite[p.~127]{Curtis:2020}. The term (I) in~\eqref{eq:Energy} represents the total specific kinetic energy, i.e., the relative specific kinetic energy supplemented with the specific kinetic energy of rotation of the non-inertial reference frame. The latter is equivalent to the specific potential energy of the centrifugal acceleration. Term~(II) is the total gravitational potential energy arising from the masses of the two main bodies. Therefore, $\flatfrac{\mathcal{E}}{m}$ represents the total mechanical energy per unit mass of the third body, and the Jacobi integral is a quantity proportional to $\flatfrac{\mathcal{E}}{m}$, which will remain constant for a third body in the CR3BP. Zero velocity curves, i.e., the location in space where the speed $v$ would go to zero, enable us to demarcate the possible regions accessible to the third body with a given potential energy. When $v$=0, the Jacobi constant is equal to the pseudo-potential $U$ in~\eqref{eq:PP}. Hence, Eq.~\eqref{eq:Energy} becomes $U(v=0)=C_{\rm J}=\Omega^2r^2 + \dfrac{2\mu_1}{r_1} + \dfrac{2\mu_2}{r_2}$. 

Figure~\ref{fig:JI}a shows examples of zero velocity curves for $C_{\rm J}\in\unit[\{150, 155, 160, 175, 185\}]{kJ/kg}$ for the CR3BP formed by Pluto--Charon and a third, smaller body (e.g., an orbiting spacecraft). Panels (b) through (f) of Figure~\ref{fig:JI} display regions inaccessible by the third body with a given Jacobi integral. For example, if the third mass has a total energy corresponding to $C_{\rm J}=\unit[150]{kJ/kg}$, it could approach $L_4$ and $L_5$, but not orbit these stable Lagrangian points within the shaded regions. These regions are sometimes referred to as `forbidden' regions.\cite{MatlabMonkey:2023}  Similarly, a body with $C_{\rm J}=\unit[155]{kJ/kg}$ could not escape the Pluto--Charon system through the $L_3$ point. A $C_{\rm J}=\unit[175]{kJ/kg}$ indicates that a satellite cannot exit the two-body system without an additional $\Delta{v}$, i.e., external increase of the speed. Finally, with a Jacobi integral $C_{\rm J}\gtrsim\unit[185]{kJ/kg}$, the third body will remain bound to either the mass $m_1$ or $m_2$ depending on its initial position. 

\section{Conclusion}\label{sec:Conclusions}

Toward the beginning of Section \ref{sec:Results}, it was remarked that a wide body of evidence supports the premise that learning and teaching complex, especially abstract, concepts can benefit from real-world examples \citep{JB93,MM93,CL96,LDMM,BAT03,DP05,RT07,JG08,RH08,King2012,RS17,BK22}. One of the most potent applications of advanced classical mechanics -- which encompasses the Lagrangian and Hamiltonian formulations -- is in orbital/celestial mechanics, as outlined in Section \ref{sec:Intro}.

Hence, in this work, we tackle the Jacobi constant from the Circular Restricted 3-Body Problem via advanced classical mechanics (the Lagrangian formulation), and thereafter provide a real-world example in the form of the Pluto--Charon system. The principal results and contributions from this work can be summarized as follows:
\begin{enumerate}
  \item The Circular Restricted 3-Body Problem (CR3BP) offers a practical illustration of the benefits of the Lagrangian formalism over the classical approach entailing Newton's laws of motion.
  \item The Pluto--Charon system offers ideal conditions to demonstrate a real-world instantiation of the CR3BP in our Solar system. 
  \item The quantitative results ensuing from our study are consistent with previous publications in the peer-reviewed literature, \onlinecite[e.g., Refs.][]{Vallado:2013, DePater:2015, Curtis:2020}. 
\end{enumerate}

\begin{acknowledgments}
We gratefully acknowledge Prof. Jean Carlos Perez at the Florida Institute of Technology for useful discussion of the Lagrangian formalism. This work was supported by the National Science Foundation under CAREER grant 2047863 to Embry-Riddle Aeronautical University.
\end{acknowledgments}

\section*{CRediT}
{\bf J\'er\'emy A. Riousset:} Conceptualization, Methodology, Software, Formal Analysis, Investigation, Resources, Data Curation, Writing - Original draft preparation, Visualization, Supervision, Project Administration, Funding Acquisition {\bf Manasvi Lingam:} Validation, Investigation, Writing - Original draft preparation.
  
\bibliographystyle{aip}
\bibliography{abbref}

\end{document}